\documentclass[11pt]{article} \topmargin=-0.5cm
\usepackage{amssymb,epsfig}
\usepackage{epstopdf}
\usepackage{color}
 \newtheorem{theorem}{Theorem}%[section]
\newtheorem{corollary}{Corollary}%[section]

\newtheorem{remark}{Remark}%[section]
\newtheorem{problem}{Problem}%[section]
\newtheorem{example}{Example}%[section]

\def\F{{\cal F}}
\def\w{\widehat}

\def\Var{{\rm Var\,}}

\def\E{{\bf E}}

\def\s{\delta}

\def\ww{\widetilde}

\def\s{\sigma}
\newcommand{\be}{\begin{equation}}
\newcommand{\ee}{\end{equation}}
\newcommand{\bd}{\begin{displaymath}}
\newcommand{\ed}{\end{displaymath}}
\newcommand{\ba}{\begin{array}{ll}}
\newcommand{\ea}{\end{array}}
\newcommand{\baa}{\begin{eqnarray}}
\newcommand{\eaa}{\end{eqnarray}}
\newcommand{\baaa}{\begin{eqnarray}}
\newcommand{\eaaa}{\end{eqnarray}}   \font\sm=cmr10
\def\ww{\tilde}

\title{ On a gap between rational annuitization price for producer and price for customer
}
 {\author{Nikolai Dokuchaev\\ \ {\sm  School of Electrical Engineering, Computing and Mathematical Sciences,}\\
{\sm  GPO Box U1987, Perth, 6845 Western Australia}\\  {\sm Tel. 61
08 92663144, Email N.Dokuchaev@curtin.edu.au}}}
\begin{document}
\maketitle
\let\thefootnote\relax\footnote{ 
This is a
pre-copy-editing, author-produced PDF of an article accepted for publication to "Journal of Revenue and Pricing Management"  following peer review. 
%?This is a pre-print of an article published in [insert journal title]. The final authenticated version is available online at: https://doi.org/[insert DOI]?.
}
\let\thefootnote\relax\footnote{This a revised version of a paper web-published on ssrn.com 
on September 21, 2011;  https://ssrn.com/abstract=1931921}
\begin{abstract} The paper studies  pricing of insurance products  focusing on the  pricing of annuities under  uncertainty. This pricing 
problem  is crucial for financial decision making  and was studied intensively;   however, many  open questions still remain.
In particular,  there is a so-called  ``annuity puzzle" related to certain  inconsistency of existing financial theory with  the empirical observations for the annuities market.  The paper suggests  a pricing method based on  the risk minimization such that
both producer and customer seek to minimize the mean square hedging error accepted as a measure of risk.
This leads to two different versions of the pricing  problem: the
selection of the  annuity price given the rate of
regular payments, and the selection of the  rate of
 payments given the annuity price.  It appears that solutions of these two problems are different.
 This can contribute to explanation for the  "annuity puzzle".
\end{abstract}
%{\itAbbreviated tittle:}
%\end{document}
\section{Introduction}
 The paper addresses the problem of pricing of  annuities under
uncertainty in the future movements of market and uncertainty in life longevity.
 More precisely, the paper  studies the problem of
pricing  for annuitization when a lump sum payment is
exchanged on the right of the periodic income payments during
certain time period.
Solution of this problem is crucial for financial decision making.
The   problem  was studied intensively; however, many  open questions still remain.
In particular,
there is so-called  "annuity puzzle" related to certain  inconsistency of existing financial theory with the empirical  observations for the annuities market.

There are two major types of annuities: life annuities, when the payments are for the rest
of annuitant's life, and  fixed term annuities, when the payments
are for a specified period of time.

For both life annuities and fixed term annuities, all  participants
of this type of a contract accept certain risk caused by uncertainty
in the  future market movements (see e.g. Yaari (1965)). For an
annuitant, it is a risk of missing better investment opportunities:
in the case of a rise of the rate of return of the investment on the
market, the agreed payments can be lower than the ones that could be
produced by the same wealth as the annuity price invested in the
market. On the other hand, the annuity seller also accepts certain
risks: in the case of a downward movement of the rate of return on the investment in
the market, the agreed payments can be higher than the
ones produced by the same wealth  invested in the market.
\index{In addition, for the life annuity, the total payments could be larger or smaller than
the original amount invested in the annuity, if  the case of
unexpectedly long or short annuitant's life. }
\par
For a more simplistic model with a deterministic rate of return of the investment
in the market, it is possible to find the fair price of a fixed term
annuity that represents the future value of money and ensures the
perfect hedge for the annuity seller. In addition, for fixed term
annuities, the market incompleteness arising from the interest rate
risk will disappear if a sufficiently deep market in government
bonds exists. Assume, for instance, that the government bonds are
available for the maturities at any lifetime of the annuity. In this
case, the price of annuity will be defined by the price of the
portfolio of the corresponding bonds. This  means that the problem
of pricing of term annuities can be  reformulated as a problem of
pricing of bonds. In fact, there is a duality between pricing
problems for bonds and term annuities; one can derive the price of
bonds via prices of annuities.
\par
For life annuities, there is an
additional  longevity risk caused by the randomness of the life
contingency for the annuitant. For the annuitant, the risk  that, should the annuitant die rather sooner than
later, he or she will receive much less money than was invested in
the case of early death; the insurance company gets to keep the
remainder of the account. For these annuities,  perfect hedging is impossible even  with a
deterministic rate of return on the investment in the market. In
addition, pricing of life annuities cannot be reformulated as a
problem of bonds pricing.

There is also a  so-called ``annuity puzzle" related to some inconsistency of empirical  data
with the economic theories  that can be
described as follows. At time of retirement many people are making
the decision to take out a lump sum of money from their retirement
account or to select an annuity payment. Economists have shown that buyers of annuities are assured more annual income for the rest of their lives. However,
historically, people by a majority opt for the lump sum even given
that this decision appears to be economically irrational.
 As a result, the annuities have a smaller market share
than can be expected from the rational investor point of view. This
was  observed by Yaari (1965); the problem was
widely studied since then.
\par
The present paper readdresses the problems of  pricing
of annuities. The goal is to develop a particular stochastic model for annuities  
where the price  is associated with certain
degree of possible risk.
One objective is to develop a pricing method  and explicit formulas for the prices.
Another objective  is to find and expose features of the
annuities that may contribute to the explanations of the "annuity puzzle".

\par
Let us describe our methodological approach.

We assume
 a market model where a potential annuitant is making a choice between a given lump
sum and the annuity. The amount
paid for the annuity by the annuitant is invested  in the market by
the annuity seller, and  that this investment  has time
variable and stochastic rate of return; for instance, it can be
investment in cash account with stochastic short term interest rate.
More precisely, we assume that this stochastic rate of return is
described  by a stochastic differential equation.

\index{  In
particular, a zero coupon bond or a stock index can be considered as
such an investment. For this model, the investment into annuity
replaces for annuitant the investment into this asset; random return
from this asset is replaced by constant cash flow from annuity.}

We consider a setting  where
both produces and customers seek to minimize the mean square hedging error as a measure of risk
(equation (\ref{error}) below).
This error represents the difference between the
cost of annuity and the future value of the total cumulate payments, and has exactly  the same value for both producer and customer. However, this  setting allows  two different versions of the pricing  problem: the
selection of the  annuity price given the rate of
regular payments, and the selection of the  rate of
 payments given the price of the annuity.  It appears that  these two problems have  different solutions.
We obtained  the corresponding pricing formulas
explicitly (Theorems \ref{ThM1}--\ref{ThM2} below).

This can  contribute  to understanding of the ``annuity puzzle"
as the following. The presence of  two different solutions implies that
there is no a fair price that minimizes risk for both the annuity
seller and the annuitant, i.e. there is no an "equilibrium"
price  (Theorem \ref{ThA} and Examples \ref{ex1}-\ref{ex2} below). Respectively, it is impossible  to find a price acceptable for  both parties because of the  perception  at least for one of the parties that a  suggested price is  unfair.  So far, behavioral aspects related to pricing anisotropy related to
risk-minimizing  has not yet been formulated as  a
reason for the ``annuity puzzle".   The existing literature addressed  different factors, namely
bequest motive,  background risk,  and  cost inflating factors
such as administrative  and marketing  expenditures and  extensive  profits. The paper  suggests one more possible factor. \par

\subsubsection*{Literature review}
A comprehensive introduction to basic principles of pricing of life annuities can  be found in  Gerber (1997) and Milevsky (1997).

The "annuity puzzle"  was  observed first by Yaari (1965) and was later studied
by a number of authors; see e.g. B\"utler and Teppa (2007)
and references therein. There were several explanations for the "annuity puzzle"  suggested in the
literature, including  a bequest motive (see e.g. Ameriks {\em et al}
(2011)), background risk (see, e.g., Horneff et al. (2009), Pang and
Warshawsky (2010)), unfair annuity prices (see e.g., Mitchell {\em et
al} (1999), Finkelstein and
Poterba (2004), Brunner and Pech 2006)), and behavioral aspects
(Brown {\em et al} (2008), Benartzi {\em et al} (2011)). A review of
related literature can be found in Schreiber and Weber (2013).
\par
 The impact of  perception of price fairness for different industries was studied in
e.g. Devlin {\em et al} (2014), Choi and Mattila (2003), Chung (2017) ,
Kienzler (2018),
Kimes {\em et al}   (2003), Xia {\em et al}  (2004), Zhan and  Lloyd  (2014).
\index{1,2,3,4,5,6,7}
A  review of studies on probabilities associated with life longevity can be found in Crawford
{\em et al} (2008).

 A review of the  mean-variance pricing can be found in  Schweizer
(2001).

\section{The problem setting and the main result}
\subsection*{Market model}
We consider a market model such that an investment of \$1 at time
$s$ generates the return $B(t,s)$ at time $t\ge s$, where $B(t,s)>0$
is random and such that $B(s,s)=1$. \index{This model is equivalent to the
model where it is possible to invest only in a zero-coupon bond with
a price $b(t)>0$ at time $t$; the corresponding value of $B(t,s)$ is
$B(t,s)=b(t)/b(s)$.}
\par
We assume that $B(T,t)$ evolves as
\baaa  &&d_tB(t,s)=B(t,s)(r(t) dt+\s(t) dw(t)),\quad t>s, \nonumber \\
&& B(s,s)=1. \label{B}\eaaa Here $w(t)$ is a standard stochastic Wiener
process; equation (\ref{B}) is a stochastic differential It\^o's equation.  The value of $\s(t)>0$ is used as the measure of the uncertainty in market movements at time $t$; larger $\s(t)$ means more uncertainty.
\par We
assume that the coefficients $r(t)$ and $\s(t)$ are random, bounded,
and  such that they are independent from the increments $w(t+\Delta
t)-w(t)$ for all $t>0$ and $\Delta t>0$. In addition, we assume that
$r(t)\ge 0$, $\s(t)\ge 0$.
\par
Consider an annuity that produces regular payments during the time
period $[0,T]$, where  $T$ is the terminal time for this annuity; $T$ is random for   life annuities, and $T$  is  non-random for fixed term annuities.
 \par
We consider the annuities with fixed payment on some time interval
$[0,T]$. Moreover, we assume that the payments are
quite frequent such that they are approximated by a continuous cash
flow with some constant density $u>0$, or the rate of payments. The value $u$ describes  the
density of the constant cash flow generated by the annuity and paid
to the annuity holder. In other words,
$u\Delta t$ is the amount of cash paid during the time period
$[t,t+\Delta t]\subset[0,T]$.
\par
Let $a$ be the wealth that has to be annuitizated.
\par
For this model, if the wealth $a$ is invested by the annuity seller,
then the wealth generated at time $T$ will be $B(T,0)a$. The current
cost $x_u(t)$ at time
$t\in[0,T]$ of the annuity payments  to the seller  is \baaa x_u(t)=\int_0^tB(t,s)u\,ds. \eaaa
\par Let $\ww x_u(t)=B(t,0)^{-1}x_u(t)$ be the discounted
current cost of the annuity payments to the seller at time
$t\in[0,T]$.  We have that \baaa \ww x_u(t)=
B(t,0)^{-1}x_u(t)=\int_0^tB(t,0)^{-1}B(t,s)u\,ds
=J(t)u, \eaaa where
$$ J(t)=\int_0^tB(s,0)^{-1}\,ds.$$
\par
 If both $T$ and
$B(t,s)$ are non-random, then the cost of serving the annuity can be perfectly
hedged either via selection of constant $u$ given $a$ or
via selection of $a$ given $u$ such that \baa B(T,0)a=x_u(T), \quad a=\ww x_u(T).\eaaa
This gives \baaa
a=J(T)u. \label{au}\eaa In the literature, it is commonly accepted
as the fair price of the annuity (see, e.g.,  Milevsky (1997)).
Respectively, in the case of non-random $B(t,s)$ and $T$, \baa
u=J(T)^{-1}a \label{ua}\eaa is the fair rate of the payments given the
annuity price $a$.

 If either $B(t,s)$ is random or $T$ is random, then the cost of serving
the annuity cannot be hedged perfectly via selection of constant
$u$. In this case, it is reasonable to calculate "optimal" risk
minimum $a$ given $u$ and"optimal" risk minimum $u$ given $a$ such
that the hedging error  $|a-\ww x_u(T)|$
 is minimal in a certain probabilistic sense. We consider minimization of
 the mean square hedging error
\baa
\E[|a-\ww
x_u(T)|^2],
\label{error}
\eaa
 where $\E$ denote the expectation.
\subsection*{Selection of  the payments and the price of annuity}
First, we consider calculation  of the  "fair"   price $a=a(u)$ of annuity that generates given constant payments $u$.
This price should be optimal with respect to risk minimization meaning
that the
discounted wealth  generated for the buyer from the initial investment  $a$
 has  the closest value to the total costs to the
seller.  For this, we state the following problem.
\begin{problem}\label{P1}
\baa
&&\hbox{Minimize}\quad\E[|a-\ww x_u(T)|^2]\quad\hbox{over}\quad
 a >0\quad \hbox{given}\quad u.
 \label{optim1} \eaa
 \end{problem}
\par
Second, we consider calculation of  "fair" payments $u=u(a)$ for the annuity sold for a given amount  of money  $a$.
These payments should be optimal with respect to risk minimization meaning
that the  their discounted total
costs $\ww x_u(T)$ to the seller of these payments
 has  the closest value to the
discount wealth generated from the initial wealth $a$.
For this, we state the following problem.
\begin{problem}\label{P2}
 \baa
&&\hbox{Minimize}\quad\E[|a-\ww x_u(T)|^2]\quad\hbox{over}\quad
 u >0\quad \hbox{given}\quad a.
 \label{optim2}
 \eaa
 \end{problem}
\par
Both problems target to minimize the same value (\ref{error}) quantifying the mean square  hedging error and characterizing the risk,  to ensure "fair" conditions of the contact  (i.e., the price
of annuity or the rate of payments). However, as is shown below, these problems have different solutions is the stochastic setting.
\par
In Theorem \ref{ThM1} below, the case of  random  $T,r,\s$ is not
excluded.
\begin{theorem}\label{ThM1} \begin{itemize}
\item[(i)] Problem \ref{P1} has a unique solution \baa \w a(u)=u \E J(T)\label{a}\eaa (It is
the actuarial present value of the total value of the benefits for
the annuitant). The second moment of the hedging error $|\w a(u)-\ww x_u(T)|$ for
this solution is \baaa R_1(u)=\E|\w a(u)-\ww x_u(T)|^2= u^2\Var J(T). \eaaa
\item[(ii)] Problem \ref{P2}     has
a unique  solution
 \baa \w u(a)=a\frac{\E
[J(T)]}{\E [J(T)^2]}.\label{u}\eaa
 The second moment of the hedging error $|a-\ww x_{\w u(a)}(T)|$ for this solution is \baaa R_2(a)=\E|a-\ww x_{\w
u(a)}(T)|^2=a^2\left[1-\frac{(\E J(T))^2}{\E [J(T)^2]}\right]. \label{R2}\eaaa
 \end{itemize}
\end{theorem}
\par
For $t\in (0,+\infty)$, set  \baa
y(t)=\E\int_0^tB(s,0)^{-1}ds,\qquad
z(t)=\E\left[\left(\int_0^tB(s,0)^{-1}ds\right)^2\right].
\label{yz}\eaa
\begin{corollary}\label{corr1} If $T$ is independent on $w(\cdot)$ then \baaa &&
\w a(u)=u \E y(T),\quad \w u(a)=a\frac{\E y(T)}{\E
z(T)}.
\eaaa
\end{corollary}

\begin{theorem}\label{ThM2}  If $T$ is non-random and given, and
if $r(t)\equiv r\ge 0$ and $\s(t)\equiv\s>0$ are given non-random constants,
then \baa &&
y(T)=\frac{1-e^{(-r+\s^2)T}}{r-\s^2},\label{y}\\
&&z(T)=\frac{2}{r-2\s^2}\left(
y(T)-\frac{1-e^{(-2r+3\s^2)T}}{2r-3\s^2}\right).\label{z}
\eaa
\end{theorem}
\par
For life annuity, the expiration time $T$ is unknown a prior and has to be modelled as a random variable; $T$ is defined  by
life time of the annuitant.  In this paper, we assume that $T$ is independent from the
process $B(\cdot,\cdot)$. If we assume  that $T$ has a probability density
function
 $\lambda(x)$, then (\ref{a}) and (\ref{u})can be rewritten as
\baa \w a(u)=u\int_0^{\infty} y(t)\lambda(t)dt,\qquad \w
u(a)=a\frac{\int_0^{\infty} y(t)\lambda(t)dt}{\int_0^{\infty}
z(t)\lambda(t)dt}\qquad \label{ur}\eaa respectively, where $y(t)$
and $z(t)$ are defined by (\ref{yz}) with non-random $T=t$.
\par
The probability density function $\lambda$ is defined by the life longevity distributions. These distributions are extremely important for the insurance industry, and they
were studied in detail; see, e.g., Tennebein and
Vanderhooof (1980) and the review of more recent studies in Crawford
{\em et al} (2008).  In practice, the  distributions of $T$ are  described by the
mortality tables; they can be used for estimation of the integrals in  (\ref{ur}).
\par Figures
\ref{fig1}-\ref{fig2}  illustrate Theorem \ref{ThM1} for the case where  $r(t)\equiv 0.05$ and $T=20$, and where $\s(t)\equiv \s$ is constant.
 Figure \ref{fig1} illustrates Theorem \ref{ThM1}(i) and shows the shape
 of dependence from $\s$ of the risk minimum annuity price $\hat a(u)$ on
$\s$ given $u=1$ defined by (\ref{a}).
 Figure \ref{fig2} illustrates Theorem \ref{ThM1}(ii) and shows the shape
 of dependence from $\s$ of the risk  minimum  payment $\hat u(a)$  given $a=1$
defined by (\ref{u}).
\begin{remark} {\rm It can be noted that maximization of the expected return for either parties is excluded from our optimization setting, because this type of maximization leads to
a non-cooperative zero-sum game where the gain of one party would lead to the  same loss for the  counterparty; in this case, an equilibrium solution is not feasible.  In our setting,  both parties target minimization  of the same value  $\E[|a-\ww x_u(T)|^2]$; this, in principle, could lead to the same equilibrium price. At least,  this equilibrium price exists in the deterministic case,
where (\ref{au}) and (\ref{ua}) give the same  solution for both Problems \ref{P1} and \ref{P2}.}
\end{remark}
\section{Impact of the presence of two different optimal solutions}
Let $\w a(u)$ be optimal (risk minimum) $a$ given $u$, and let $\w
u(a)$ be optimal $u$ given $a$; they are defined by  (\ref{a}) and  (\ref{u}) respectively.   For the case of deterministic model
with non-random $T$ and $B(t,s)$, there exists $k>0$ such that $\w
a(u)=ku$  and $\w u(a)=k^{-1}a$; in other words, \baaa\frac{\w
u(a)}{a}=\frac{u}{\w a(u)}. \eaaa
 It does not hold for the stochastic model.
\begin{theorem}\label{ThA} If $\Var J(T)>0$,  then \baa \frac{\w u(a)}{a}<\frac{u}{\w a(u)}.
\label{ineq}\eaa
\end{theorem}
\par
It can be noted that  if either $B(t,s)$ is random and $T$ is non-random, or $T$ is
random and $B(t,s)$ is non-random, then  $\Var J(T)>0$.
\par
 According to Theorem \ref{ThA},  the relative price of annuity is
larger for the setting of Problem \ref{P1} then for the setting of Problem \ref{P2}.
Respectively,  the rate  return  on the annuity investment  is
smaller for the setting of Problem \ref{P1} then for the setting of Problem \ref{P2}.

\index{In this sense, the  the mean-variance pricing is  anisotropic
with respect to the direction of Anisotropy of mean-variance
pricing.}
\par
Inequality (\ref{ineq}) is illustrated by Figures
\ref{fig3}-\ref{fig4}.
 Figure \ref{fig3} shows the shape
 of dependence from $\s$ of the difference $\frac{\w u(a)}{a}-\frac{u}{\w a(u)}$.
It  can  be seen from this figure   that the value of
$\frac{\w u(a)}{a}-\frac{u}{\w
 a(u)}$ vanishes for larger $\s$. This can be explained by the fact
 that $\w u(a)$ with given $a$ decreases when $\s$  increases and
 that  $\w a(u)$ with given $u$ increases when $\s$  increases.  Figure
 \ref{fig4}  shows
 the dependence from $\s$ for the ratio $\frac{\w u(a)}{a}\Bigl/\frac{u}{\w a(u)}$. It can be seen that this value is separated from 1 even for
 larger $\s$.

 For Figures
\ref{fig3}-\ref{fig4}, we assumed again that $r=0.05$ and $T=20$,
similarly to Figures \ref{fig1}-\ref{fig2}. \subsection*{Economic
interpretation and annuity puzzle} Theorem \ref{ThA} implies that,
even in our simple model, there is no a fair
price that is optimal for the annuity seller as well as for
annuitant, or an "equilibrium" price.
  As can be seen from Figures
\ref{fig3}-\ref{fig4}, there is a spread between risk minimum prices
calculated with  different selection of what is fixed initially, $u$
or $a$. This may contribute to the explanations of the   so-called
"annuity puzzle": a statistically confirmed fact that people does
not choose annuity and opt for the lump sum even given that this
decision appears to contradict standard theoretical analysis; see
the references provided in Section 1.

Let us consider the following  Examples \ref{ex1}--\ref{ex2}. For these examples, we assume, in the framework of
our model, that $\s(t)\equiv 0.2$ and $r(t)\equiv 0.05$.
\begin{example}\label{ex1}{\rm
Assume that a potential
annuitant wishes to convert her life savings, $a=\$100,000$, into a term annuity for 20
years.  A potential seller calculates the optimal rate of payments as $\w
u(a)=\$4,206$  solving Problem \ref{P2} via (\ref{u}),
and will consider a larger rate of payments to be unfair. However, the
potential annuitant calculates that this rate of payments  would
require optimal investment of $\w a(\$4,206)=\$76,242$ only,
solving Problem \ref{P1} via (\ref{a}). This means that there is a spread between rational  annuitization prices for the customer
and producer. This may lead  to a disagreement about the conditions of
the contract and the failure of the deal.
}\end{example}
\begin{example}\label{ex2}{\rm
Similarly, assume, that, for the same model, a seller
offers  an annuity with the payment rate $u=\$5,000$ for 20 years.
A potential annuitant  calculates   the optimal price $\w
a(u)=\$90,635$ solving Problem \ref{P1} via (\ref{a}),
and will consider a larger price to be unfair. However, the seller
calculates  that the investment of this size  should ensure the rate
of payments $\w u(\$90,635)= \$3,812$ only, solving Problem \ref{P2} via (\ref{u}). Again, this may lead to a
disagreement about the price and the failure of the deal.
}\end{example}

These spreads between prices can lead
 to a market failure, when there is no sufficient supply for the
prices acceptable for the customer. \index{This can be illustrated by
the fact that, currently,  there is only one
company that offers  life annuities
 in Australia ({\em Challenger Limited}).}

\begin{remark}\label{remE12}{\rm We presume  that it is more typical
that an annuity buyer has a certain fixed amount of pension money to invest and shops for a fair annuity rate found as the solution of Problem \ref{P2}
rather than target  a preselected rate of payments  and selecting a fair lump sum that buys it
according to Problem \ref{P1}.
\index{%By Theorem \ref{ThA}, this means that
%it is the annuitant  who  perceives the offer from the insurer as unfair.  }
Otherwise, the situations described in Examples \ref{ex1} and \ref{ex2} will be reversed.
}\end{remark}

  \index{Let us assume the following
model: a potential annuitant is making a choice between a given lump
sum and the annuity.}

\index{ If there is a market competition among annuities sellers,
then this should lead to the pricing rule (\ref{a}).}
\section{Proofs}
The proofs below are rather technical; they use some basic facts from stochastic
analysis, namely formulae for expectations of the It\^o's integrals; see, e.g., a review in Dokuchaev (2007).

 \par {\em Proof of Theorem \ref{ThM1}.} It follows from ({\ref{B})
that \baa
B(t,s)=\exp\left[\int_s^tr(\tau)d\tau-\frac{1}{2}\int_s^t\s(\tau)^2d\tau+\int_s^t\s(\tau)dw(\tau)\right].
\label{B1}\eaa
\par
(i) The function $V(a,u)=\E|\ww x_u(T)-a|^2$ can be rewritten as
\baaa &&V(a,u)=u^2\E [J(T)^2]-2au\E J(T) + a^2. \eaaa Hence
\baa V(a,u)=(a -u\E J(T))^2+ R_1(u), \label{va}\eaa where
\baaa R_1(u)= u^2\E [J(T)^2]-u^2(\E J(T))^2=u^2\Var  J(T) . \eaaa
For a given
$a$, the function $V(a,u)$ has the only minimum at $\w a(u)$ given by (\ref{a}), and $V(\w a(u),u)=R_1(u)$.
\par
(ii) Similarly to (\ref{va}), the function $V(a,u)=\E|\ww
x_u(T)-a|^2$ can be rewritten as  \baaa &&V(a,u)=
\left(u (\E
[J(T)^2])^{1/2}-\frac{a\E J(T)}{(\E [J(T)^2])^{1/2}}\right)^2+ R_2(a), \eaaa where $R_2(a)$ is defined by (\ref{R2}).
For a given
$a$, the function $V(a,u)$ has the only minimum at $\w u(a)$ given by (\ref{u}), and $V(u,\w u(a))=R_2(a)$. This
completes the proof of statement (ii) and Theorem \ref{ThM1}.
 \par {\em Proof of Corollary \ref{corr1}.}
  Since $T$ is
independent from $w(\cdot)$, it follows that \baa \E J(T)=
\E[\E\{J(T)|T\}]=\E y(T),\quad  \E J(T)^2=
\E[\E\{J(T)^2|T\}]=\E z(T).\label{ucon}\eaa The the statement of the Corollary follows.
  \par {\em Proof of Theorem \ref{ThM2}.} Let $\rho=r-\s^2/2$.
 By (\ref{B1}), we obtain that, for non-random $T$, \baaa
y(T)=\E\left[\int_0^TB(t,0)^{-1}dt\right] =\E\left(\int_0^T e^{-\rho
t-\s w(t)}dt\right)=\E\int_0^T e^{-\rho t-\s w(t)}dt\nonumber \\=\E\int_0^T
e^{-\rho t +\s^2t/2}= \E\int_0^T  e^{-r t +\s^2t}=
\frac{1-e^{(-r+\s^2)T}}{r-\s^2}.\eaaa  Then ({\ref{y}) follows.
Further, we have that, for non-random $T$, \baaa
z(T)=\E\left[\left(\int_0^TB(t,0)^{-1}dt\right)^2\right]
=\E\left(\int_0^T e^{-\rho t-\s w(t)}dt\right)^2\nonumber\\=\E\int_0^T\int_0^T
e^{-\rho(t+s)-\s [w(t)+w(s)]}dtds=2\E\int_0^Tdt\int_0^t
e^{-\rho(t+s)-\s [w(t)+w(s)]}ds\nonumber\\ =2\E\int_0^Tdt\int_0^t
e^{-\rho(t+s)-\s w(s) }\E\{ e^{-\s w(t)}|\F_s\}ds.\eaaa
 For $s<t$,
we have $\E\{e^{-\s(w(t)}|\F_s\}=e^{-\s w(s)}e^{\s^2(t-s)/2}$ and
\baaa z(T)&=&2\E\int_0^Tdt\int_0^t  e^{-\rho(t+s)-\s w(s) }e^{-\s
w(s)}e^{\s^2(t-s)/2} ds\nonumber\\ &=&2\int_0^Tdt\int_0^t
e^{-\rho(t+s)}e^{2\s^2s}e^{\s^2(t-s)/2} ds. \eaaa Taking the last
integral, we obtain that \baaa &&z(T)= 2\int_0^Tdt\int_0^t
e^{t(-\rho+\s^2/2)}e^{s(-\rho+2\s^2-\s^2/2)}
ds\nonumber\\&&=2\int_0^Tdt\int_0^t
e^{(-r+\s^2)t}e^{s(-r+\s^2/2+2\s^2-\s^2/2)} ds
=2\int_0^Tdte^{(-r+\s^2)t}\int_0^t e^{s(-r+2\s^2)}  ds \nonumber\\&&=2\int_0^Te^{(-r+\s^2)t}\frac{1-e^{(-r+2\s^2)t}}{r-2\s^2}dt.
\eaaa
Hence
\baaa
 z(T)&=&\frac{2}{r-2\s^2}\left[ \frac{1-e^{(-r+\s^2)T}}{r-\s^2}-\frac{1-e^{(-2r+3\s^2)T}}{2r-3\s^2}\right]\nonumber\\
 &=&\frac{2}{r-2\s^2}\left(
y(T)-\frac{1-e^{(-2r+3\s^2)T}}{2r-3\s^2}\right).
 \eaaa
 Then ({\ref{z}) follows.
 This completes the proof of Theorem \ref{ThM2}.
 $\Box$

 \par {\em Proof of Theorem \ref{ThA}.} By Theorem \ref{ThM1} and by independence of $T$ from $B(t,s)$,
 it follows that
\baaa \w a(u)=u\E y(T)=u\E J(T),\quad  \w u(a)=a\frac{\E y(T)}{\E
z(T)}=a\frac{\E J(T)}{\E (J(T)^2)}.\label{wau}\eaaa Clearly, $\E
[J(T)^2]\ge  (\E J(T))^2$, and  if $\E J(T)^2= (\E J(T))^2$ then
$\Var J(T)=0$. It follows that \baaa \frac{\w u(a)}{a}=\frac{\E
J(T)}{\E( J(T)^2)}< \frac{\E J(T)}{(\E [J(T)])^2}= \frac{1}{\E
J(T)}=\frac{u}{\w a(u)}.\eaaa
 This completes the proof of Theorem \ref{ThA}. $\Box$
\section{Discussion}
The paper analyzes pricing method  based on
minimization of the risk associated
with annuitization and caused by uncertainties in future bank rates
and life longevity. The paper points out on the existence of two different  versions of this problem:  the  risk
minimizing selection of the annuity price given the size of the regular
payments, and  the risk minimizing selection of the size of the
regular payments given the amount originally invested into the
annuity.  We found that, under uncertainty, these two problems have different solutions, even
given that they both  minimize the same value (\ref{error}). So far, this feature has been overlooked in the existing literature. The gap between these two solutions for
the customer and producer may  contribute to the so-called annuity
puzzle. In particular,  customers might regard the (risk-minimizing)
payment stream $\w u$, offered by a life insurance company for a given price $a$, as being too low.
Similarly,  producers  might regard the risk-minimizing payment
 $\w a$ offered by the customer  as being too low.
 This adds to the set of arguments
identifying  reasons for lacking annuity demand and why the annuity market is thin.

The approach suggested in the paper  allows further development in several directions.

First, it  would be interesting to investigate if this feature holds
for more advanced  models  with  multiple investments  assets and transaction costs. The particular  model presented in this paper is relatively simple and  yet it captures the presence of the gap between rational prices for the producer and customer.   Our conjecture is that this feature will hold for other market models and for other risk measures.

Second, it would be interesting to investigate discrete time models and impact of time  discretization on the pricing formulae.

Finally, it would be interesting to reverse the pricing formulae for the sake of inference  of the market parameters presented in these pricing  formulae from the  observed annuities prices.
This would follow a classical approach where the inference of the volatility of the stock prices 
is reduced to calculation of the implied volatility from the inverted  Black-Scholes pricing formula applied to stock option prices. 
Examples of extensions of this approach  can be found in
Dokuchaev (2018)  and Hin and Dokuchaev (2016a,b).  So far, the implied market parameters for the annuities market have not been considered in the literature.

%We leave this  for the future research.

\index{ \subsection*{Acknowledgments}  This work  was supported by ARC grant of Australia DP120100928 to the author.}
%\section*{ References}$\hphantom{xx}$

\newpage
 \begin{figure}[ht]
\centerline{\psfig{figure=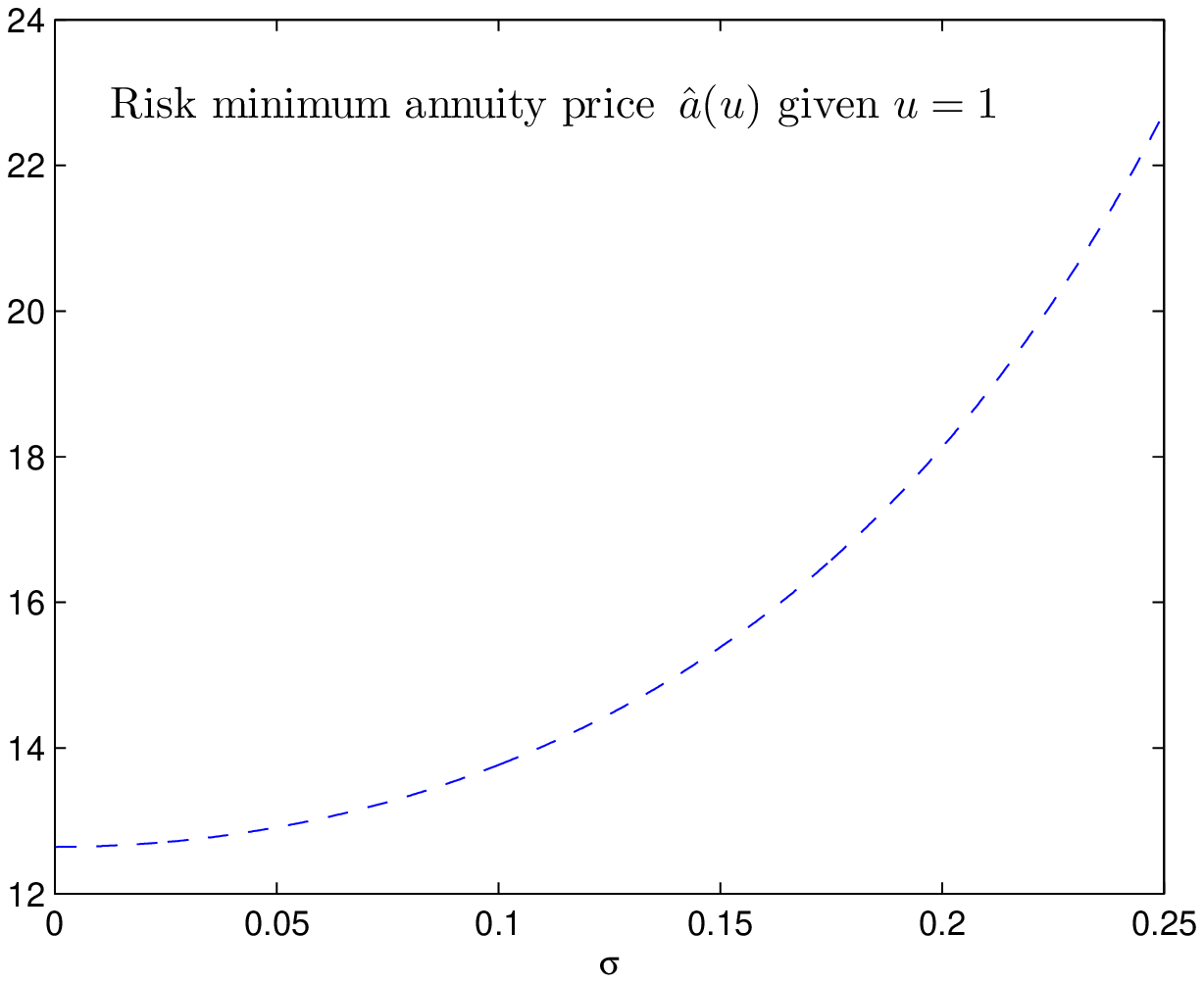,height=8.5cm}} \caption[]{ The
profile of dependence from $\s$ of risk minimum annuity price $\hat
a(u)$ given $u=1$ defined by (\ref{u}) with $r=0.05$ and $T=20$. }
\vspace{0cm}\label{fig1}\end{figure}
\newpage
 \begin{figure}[ht]
\centerline{\psfig{figure=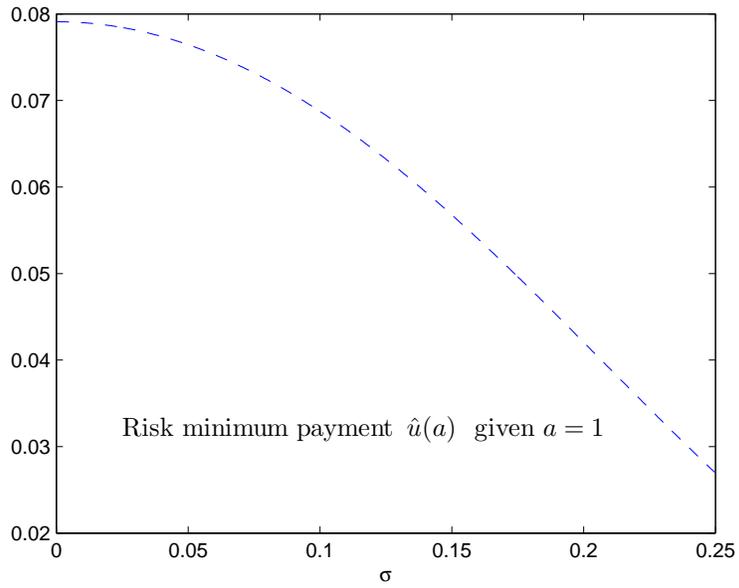,height=8.5cm}} \caption[]{ The
profile of dependence from $\s$ of the risk  minimum  payment $\hat
u(a)$ given $a=1$   given $u=1$' defined by (\ref{u}) with $r=0.05$
and $T=20$. } \vspace{0cm}\label{fig2}\end{figure}
\newpage  \begin{figure}[ht]
\centerline{\psfig{figure=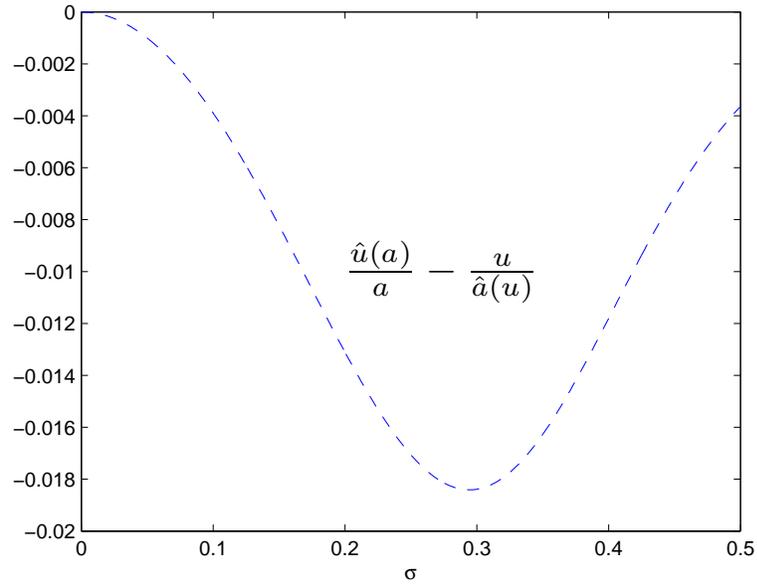,height=8.5cm}} \caption[]{ The
profile of dependence from $\s$ of  the difference $\frac{\w
u(a)}{a}-\frac{u}{\w a(u)}$ with $r=0.05$ and $T=20$. }
\vspace{0cm}\label{fig3}\end{figure}
\newpage
 \begin{figure}[ht]
\centerline{\psfig{figure=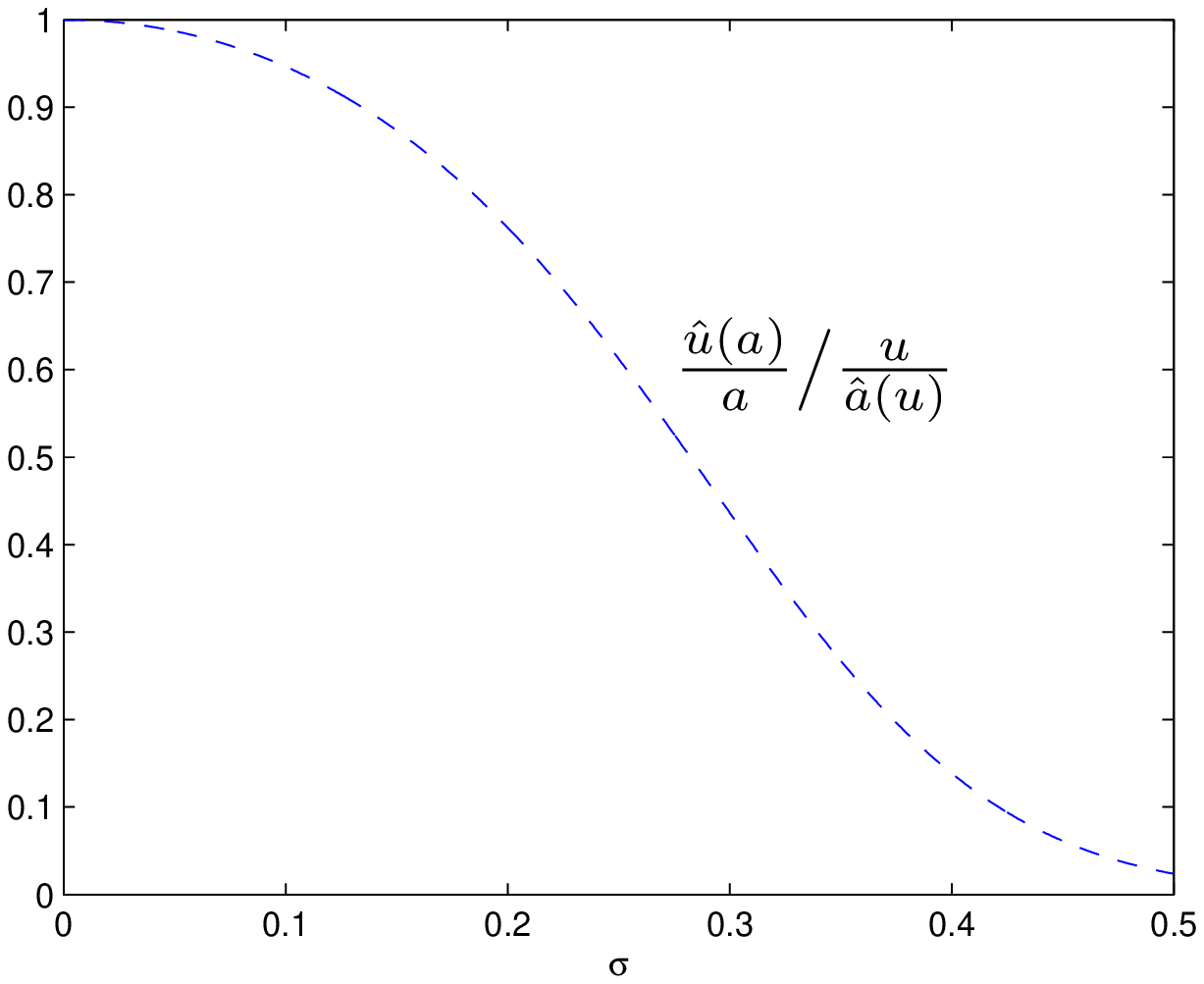,height=8.5cm}} \caption[]{ The
profile of dependence from $\s$ of  the ratio $\frac{\w
u(a)}{a}\bigl/\frac{u}{\w a(u)}$ with $r=0.05$ and $T=20$. }
\vspace{0cm}\label{fig4}\end{figure}

\end{document}